%% file: DY-PDF2ISR-2026-06-09.tex
\documentclass[11pt]{article} 
\usepackage{mystyle-new}
\usepackage{authblk}
\usepackage{bm}
\usepackage[T1]{fontenc}
\usepackage{mathtools}
\usepackage{epsfig,amsmath} 
\usepackage{hyperref}
\usepackage{color}
\usepackage{graphicx}
\usepackage{subcaption}
\usepackage{orcidlink}
\usepackage{xspace}
\usepackage{cprotect}
\definecolor{red}{rgb}{1,0,0}
\def\lesssim{\ \hbox{\raise 2pt \hbox{$<$} \kern -13pt
                     \lower 3pt \hbox{$\sim$}}\ }
\def\greatersim{\ \hbox{\raise 2pt \hbox{$>$} \kern -13pt
                     \lower 3pt \hbox{$\sim$}}\ }

\def\lsim{\mathrel{\rlap{\lower4pt\hbox{\hskip1pt$\sim$}}
    \raise1pt\hbox{$<$}}}                
\def\gsim{\mathrel{\rlap{\lower4pt\hbox{\hskip1pt$\sim$}}
    \raise1pt\hbox{$>$}}}                

\def\cascade{{\sc Cascade}}
\def\pythia{{\sc Pythia}}
\def\herwig{{\sc Herwig}}

\def\desepsf(#1 width #2){\epsfxsize=#2 \epsfbox{#1}}
\def\kt{\ensuremath{k_{\rm T}}}

\def\pt{\ensuremath{p_{\rm T}}}
\def\PZ{\ensuremath{Z}}

\def\qt{\ensuremath{q_{\rm T}}}

\def\zM{\ensuremath{z_{\rm M}}}
\def\ptll{\ensuremath{p_{\rm T}(\ell\ell)}}
\def\qs{\ensuremath{q_{\rm s}}}
\def\sqrts{\ensuremath{\sqrt{s}}}

\newcommand{\mdy}{\ensuremath{m_{\text{DY}}}}

\newcommand{\PB}{PB}
\newcommand{\PBset}{{PB-NLO-2018}}

\newcommand{\MCatNLO}{{\sc MadGraph5\_aMC@NLO}}

\def\pdfisr{{\scshape{Pdf2Isr}}\xspace}
\def\pythiaPB{{\scshape{Pythia8-Pdf2Isr}}\xspace}

\newcommand{\as}{\ensuremath{\alpha_s}}

\newcommand{\GeV}{\text{GeV}}
\newcommand{\TeV}{\text{TeV} }

\newenvironment{tolerant}[1]{\par\tolerance=#1\relax}{ \par }
\usepackage{amsmath,bm}
\usepackage{lineno}

\usepackage{cite,./mcite}
\usepackage{tikz}
\usepackage[symbol]{footmisc}

\providecommand{\DOI}[1]{\href{http://dx.doi.org/#1}}

\begin{document}

\title{Non-perturbative effects and soft-gluon dynamics in low-\pt\ Drell--Yan production}

\author[1]{D.~Subotić \thanks{dusan.sub@ucg.ac.me}\orcidlink{0009-0000-7069-5651}}
\affil[1]{Faculty of Science and Mathematics, University of Montenegro, Podgorica, Montenegro}
\author[2,3]{H.~Jung \thanks{hannes.jung@desy.de}\orcidlink{0000-0002-2964-9845}}
\affil[2]{University of Antwerp, Antwerp, Belgium}
\affil[3]{II. Institut f\"ur Theoretische Physik, Universit\"at Hamburg,  Hamburg, Germany}
\author[4]{A. V. Kotikov \thanks{kotikov@theor.jinr.ru}\orcidlink{0000-0002-1408-2735}}
\author[1]{N.~Rai\v cevi\' c\thanks{natasar@ucg.ac.me}\orcidlink{0000-0002-2386-2290}}
\affil[4]{Bogoliubov Laboratory of Theoretical Physics, Joint Institute for Nuclear Research, Dubna, Russia}

\date{}
\begin{titlepage} 
\maketitle
\thispagestyle{empty}
\begin{flushright}
\end{flushright}

\abstract{
The transverse-momentum spectrum of Drell--Yan lepton pairs at small \ptll\ probes non-perturbative QCD effects, including intrinsic partonic transverse momentum and initial-state soft-gluon radiation.
The novel \pythiaPB\ approach, which implements the Parton Branching (\PB ) framework in \pythia\ while employing the same evolution ingredients,  is used to study this spectrum in the low-\ptll\ region.
This framework is particularly well suited for such investigations, as it allows for a systematic treatment of the dominant non-perturbative effects and their interplay.

We examine in detail the implementation of primordial transverse momentum in \pythiaPB . With an improved recoil treatment, we determine the width of the intrinsic-\kt\ distribution and find a \sqrts\ dependence that is steeper than in \PB , but much flatter than in \pythia . We identify the origin of these differences. 

Furthermore, different treatments of the strong coupling at low scales are investigated and confronted with available experimental data. We find that, at the highest centre-of-mass energy considered, the low-\ptll\  Drell--Yan spectrum becomes sensitive to the transition between the perturbative and non-perturbative regimes.
}

\end{titlepage}

\section{Introduction}
The production of Drell--Yan (DY) lepton pairs at hadron colliders is a benchmark process
for testing QCD factorization and parton evolution. The region
of small transverse momentum of the lepton pair, \ptll , is of particular interest, as it probes the
transition between perturbative and non-perturbative QCD dynamics.

At low \ptll , the observed transverse momentum does not originate from a single
source. Instead, it arises from the combined action of two physically distinct mechanisms.
The first is the \emph{intrinsic transverse momentum} of partons inside the proton,
generated at a low, non-perturbative input scale and reflecting genuine proton structure.
The second is \emph{soft-gluon radiation} from initial-state parton evolution,
which builds up transverse momentum dynamically through multiple emissions governed by
the behaviour of \as\ at small scales.

A realistic description of the low-\pt\ DY spectrum therefore requires the simultaneous
inclusion of both effects. Intrinsic-\kt\ alone provides only a baseline smearing, while 
soft-gluon radiation alone cannot account for the observed width and shape of the
spectrum at the lowest transverse momenta. The measured DY distributions constrain the
interplay of these two contributions rather than either of them in isolation.

\begin{tolerant}{9000}
This interplay becomes particularly transparent when comparing
transverse-momentum-dependent (TMD) approaches\cite{Angeles-Martinez:2015sea,Boussarie:2023aa} based on the Parton Branching (PB)~\cite{Hautmann:2017fcj,Hautmann:2017xtx}
method, with initial-state parton showers,
such as \pdfisr~\cite{Jung:2025mtd}, constructed to be consistent with collinear parton density functions (PDFs). In \PB-TMD calculations, the intrinsic transverse momentum is introduced
explicitly via a non-perturbative TMD input at a low scale, while soft gluons are treated
dynamically during evolution. In the \pdfisr\ approach, the same physics is realised within
a Monte Carlo parton shower that follows the PDF evolution consistently and allows direct
access to the treatment of soft-gluon emissions and the infrared behaviour of \as .
Once soft-gluon radiation is treated consistently in the initial state, a universal
intrinsic-\kt\ width describes DY data over a wide range of centre-of-mass energies, as shown explicitly for the case of \PB -TMDs~\cite{Bubanja:2024puv,Bubanja:2023nrd}.

The conventional  \PB\ approach is based on the \PB -TMD method implemented in the 
\cascade\ Monte Carlo generator~\cite{Baranov:2021uol}, while \pythiaPB implements the 
\PB-based initial-state shower within \pythia\,8~\cite{Bierlich:2022pfr,Sjostrand:2014zea}.

In this paper we study in detail the behaviour of \pythiaPB\ in terms of the intrinsic-\kt\ distribution, the role of soft gluons, and especially the treatment of \as\ at small scales, in comparison with the \PB\ method. The predictions of the non-perturbative effects incorporated within \pythiaPB are validated through comparison with experimental measurements over a wide range of collision energies.

The novel aspects of this work are: (i) a recoil prescription in \pythiaPB\ that preserves the intrinsic transverse momentum of hard initiators and allows a relation to \PB -TMD inputs; (ii) a quantitative demonstration that the  intrinsic-\kt\ width required to describe DY data at different energies does not coincide with the \PB\ result; and (iii) a systematic investigation of the sensitivity of low-\ptll\ DY spectra to the infrared behaviour of \as\ within an ISR-consistent Monte Carlo framework.
\end{tolerant}

\section{\pythiaPB\ and  DY pair production}
The initial-state parton shower \pythiaPB~\cite{Jung:2025mtd} was developed to follow the evolution of the underlying collinear parton densities and to reproduce TMD parton densities compatible with those obtained in the PB method.

Ref.~\cite{Jung:2025mtd} shows that using the same ordering conditions, splitting-function order, and kinematic constraints as in the PDFs yields a parton shower that reproduces the same effective TMD as obtained from the TMD-PDF. For the benchmark test, the \PBset\ collinear and TMD parton densities~\cite{Martinez:2018jxt} were used.

The DGLAP evolution equation can be expressed in terms of the Sudakov form factor $\Delta^S_a( \mu^2 , \mu'^2 )$ as:
\begin{multline}
\label{EvolEqSudakov}
  {x f}_a(x,\mu^2)  =  \Delta^S_a (  \mu^2, \mu^2_0  ) \  {x f}_a(x,\mu^2_0)  \\
+ \sum_b
\int^{\mu^2}_{\mu^2_0} 
{\frac{d q'^2 } 
{q'^2} } 
{
{\Delta^S_a (  \mu^2  ,q'^2 ) }
}
\int_x^{\zM} {dz} \;
 P^{(R)}_{ab}  \left(\as ,z \right) \frac{x}{z}
\;{f}_b\left({\frac{x}{z}},
q'^2\right) .
\end{multline}
Here, $q'$ denotes the evolution scale,  $z$ is the ratio of the longitudinal momenta of the involved partons, and $P^{(R)}_{ab}$ are the regularized DGLAP splitting functions describing the splitting of parton $b$ into parton $a$. The parameter \zM\ separates resolvable emissions from the unresolved soft region. The Sudakov form factor is given by:
\begin{equation}
\label{sud-def}
\Delta^S_a ( \mu^2 , \mu'^2 ) = 
\exp \left(  -  \sum_b  
\int^{\mu^2}_{\mu'^2} 
{\frac{d { q'}^{ 2} } 
 {{q'}^{2}} } 
 \int_0^{\zM} dz  
\ z  P_{ba}^{(R)} \left(\as ,z \right)
\right) 
 .   
\end{equation}

The reversed indices in the previous equation reflect the fact that the Sudakov factor for parton $a$ sums over all unresolved branchings of the evolving parton.
In backward evolution, the daughter parton carries longitudinal momentum fraction $x$, while the parent parton carries $x'=x/z$. This leads to the backward-evolution Sudakov form factor:
\begin{equation}
\Delta_{bw} (\mu^2, \mu^2_{i-1}) = \exp \left( - \sum_b \int_{\mu_{i-1}^2}^{\mu^2}  \frac{d q^{\prime\,2}}{q^{\prime\,2}}\int_x^{\zM} dz P^{(R)}_{ab} \left(\as ,z \right) \frac{x' f_b(x',q'^2)}{xf_a(x,q'^2)} \right) .
\label{Suda}
\end{equation}

In order to make the parton shower in \pythiaPB consistent with the \PBset\ PDFs, angular ordering is applied, along with next-to-leading order (NLO) DGLAP splitting functions and maximum  longitudinal momentum fraction $\zM \to 1$.

The DY lepton-pair events are generated at NLO with
\MCatNLO~\cite{Alwall:2014hca},  applying \herwig~6 subtraction terms,  since the parton shower follows angular ordering, as described in Ref.~\cite{Yang:2022qgk}. 
The collinear \PBset~Set2  NLO parton densities 
are used.
The procedure of matching NLO with \PB\ has been detailed in Refs.~\cite{Bubanja:2023nrd,BermudezMartinez:2020tys,Martinez:2019mwt,Yang:2022qgk}.

A good description of \ptll\ is obtained in \PB\ using  \PBset~Set2~\cite{Bubanja:2023nrd,BermudezMartinez:2020tys,Martinez:2019mwt}. 
A comparison between \PB\ and \pythiaPB\ was presented in Ref.~\cite{Subotic:20254O}, where good agreement was found for the dilepton transverse-momentum distribution, up to residual differences associated with the event-level shower kinematics.

Here we investigate in detail the two contributions, the intrinsic-\kt\ distribution and the contribution of soft gluons.
In \PBset~Set2, \as\ is evaluated at a scale given by the transverse momentum of the emitted parton, which is $\qt = (1-z) q'$, and the non-perturbative region at low \qt\ is treated by freezing $\alpha_s$ below the infrared cutoff.
Since \as\ enters both the Sudakov form factor for evolution and the corresponding one for the parton shower, the non-perturbative region of \as\ becomes important and affects the \ptll\ distribution of DY pairs.

\subsection{%
  Intrinsic-\texorpdfstring{$\bm{\kt}$}{k\_T} distributions in \PB\ and \pythiaPB
}
\label{subsec:primordialKT_PB}

In \PB-TMD distributions, the intrinsic-$\kt$ distribution is introduced at the starting
scale $\mu_0$ of the evolution through a TMD distribution
${\cal A}_a(x,{\bf k},\mu_0^2)$, with ${\bf k}$ being the two-dimensional transverse-momentum vector and $\kt = |{\bf k}|$. This distribution is parametrised in terms of a
collinear parton density at the starting scale, convoluted with an intrinsic-$\kt$
distribution described as a Gaussian of width $\sigma$:
\begin{equation}
\label{TMD_A0}
{\cal A}_{a}(x,{\bf k},\mu_0^2)
= f_{0,a}(x,\mu_0^2)\;
\exp\!\left(-\frac{\kt^2}{2\sigma^2}\right)\big/ (2\pi\sigma^2)\;.
\end{equation}
The Gaussian width $\sigma$ of the intrinsic-$\kt$ distribution is related to the
commonly used root-mean-square intrinsic-transverse-momentum parameter $\qs$ via
$\qs=\sqrt{2}\,\sigma$. In \pythia, the primordial transverse momentum (intrinsic $\kt$)
of each hard-scattering initiator is generated from a two-dimensional Gaussian
distribution, in close analogy to the \PB\ input. However, there are differences in the
treatment of parton momenta in the recoil-sharing step.

\begin{figure}[htbp]
\centering
\includegraphics[width=0.4\linewidth]{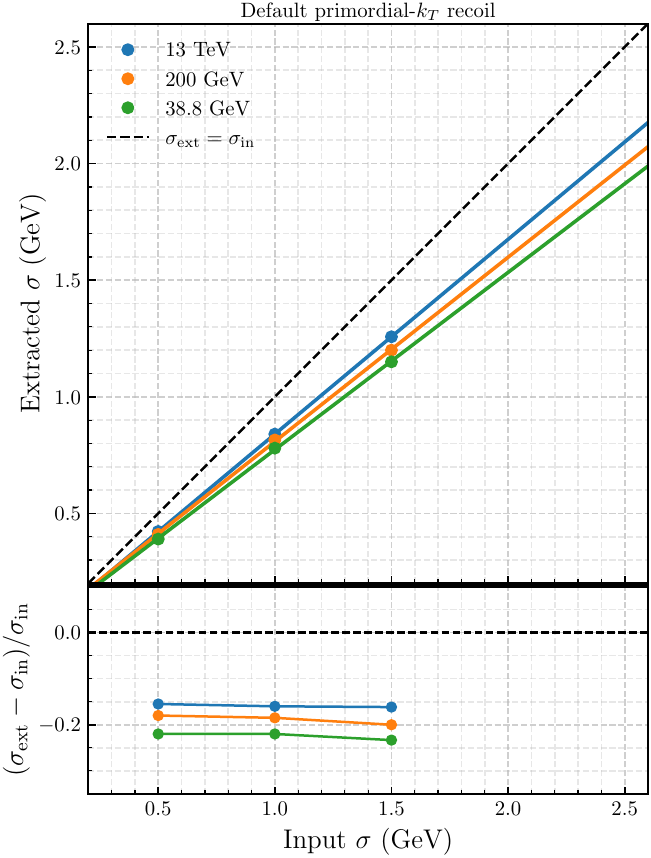}
\includegraphics[width=0.4\linewidth]{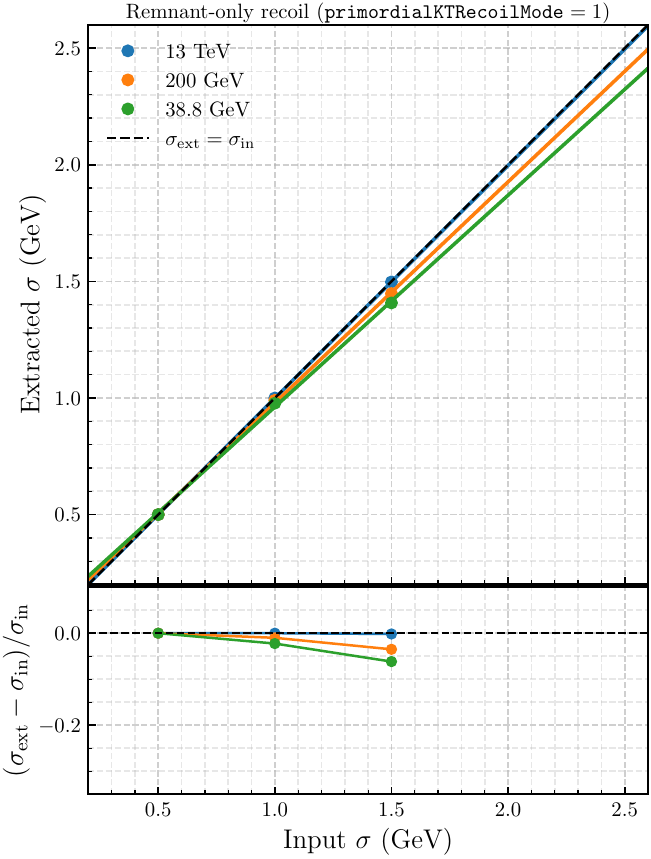}
\cprotect\caption {\small Comparison of input  $\sigma$ to the extracted one as a function of the centre-of-mass energy $\sqrts$ obtained from \pythia -default (left) and the modified recoil treatment (right).}
\label{fig:pythia-kt}
\end{figure}
In the standard \textsc{Pythia}\,8 approach, a recoil-sharing step enforces exact
conservation of the transverse momentum of the beam system. This compensation
redistributes the net transverse momentum among all beam-related partons, including the
hard initiators and beam remnants, and leads to an effective reduction of the observed
intrinsic-$\kt$ width. As a consequence, the  Gaussian width extracted from the
incoming partons or from colour-singlet transverse-momentum spectra does not coincide
with the input parameter $\sigma$. This effect is shown in the left panel of
Fig.~\ref{fig:pythia-kt}: a clear deviation of the extracted $\sigma$ from the input
$\sigma$ is observed, with a mild $\sqrts$ dependence.

While this shift is not a principal issue, it makes the comparison of the intrinsic-$\kt$
width with other approaches difficult. For comparisons with \PB-TMD calculations, it is
advantageous to preserve a direct and transparent relation between the input
non-perturbative transverse-momentum distribution and the extracted transverse momentum.
We therefore introduce a modified recoil prescription in \pythiaPB, in which the intrinsic
transverse momentum of the hard initiating partons is kept  as generated, while
the full recoil required by transverse-momentum conservation,
\begin{equation}
\sum_{\text{initiators}} {\bf k}_\text{T}\;+\;
\sum_{\text{remnants}} {\bf k}_\text{T}
= {\bf 0}\,,
\end{equation}
is absorbed exclusively by the beam remnants.

\begin{table}[!b]
\centering
\caption{Comparison of primordial-$\kt$ recoil schemes in \textsc{Pythia}\,8 and
their relation to PB-TMD modelling.}
\label{tab:recoil_comparison}
\begin{tabular}{lcc}
\hline
 & Default \textsc{Pythia}\,8 & Modified recoil scheme \\
\hline
Initiator $p_x,p_y$ distribution
 & Gaussian, rescaled
 & Gaussian, preserved \\
Effective Gaussian width
 & $\sigma_{\text{eff}} < \sigma$
 & $\sigma_{\text{eff}} = \sigma$ \\
Recoil carrier
 & Initiators + remnants
 & Remnants only \\
Beam-\pt\ conservation
 & Exact
 & Exact \\
Mapping to PB-TMD input
 & Indirect
 & Direct ($\sigma_{\text{PB}} = \sigma$) \\
Intended usage
 & General-purpose simulation
 & Theory / TMD comparison \\
\hline
\end{tabular}
\end{table}

Operationally, this is achieved by modifying the recoil-sharing step in
\textsc{Pythia}\,8, such that the compensating transverse momentum,
\(
\Delta{\bf k}_\text{T} = -\sum_{\text{initiators}} {\bf k}_\text{T}
\),
is distributed only among remnant partons, weighted by their standard recoil factors,
while leaving transverse momenta of the initiators unchanged. In the \pythiaPB\ framework, this  approach is labelled
{\small\texttt{BeamRemnants:primordialKTRecoilMode}}. With this treatment, the correlation
between the extracted $\sigma$ and the input $\sigma$ is significantly improved, as shown in the
right panel of Fig.~\ref{fig:pythia-kt}. The remaining small \sqrts\ dependence presumably arises from energy-momentum conservation effects when a physical process is considered. A summary of the two recoil strategies is given
in Tab.~\ref{tab:recoil_comparison}. The modified recoil prescription affects only the
treatment of primordial transverse momentum and its compensation within the beam
remnants. It does not rely on properties specific to Drell--Yan production and can
therefore be applied equally to other final states.

\begin{figure}[!t]
\centering
\includegraphics[width=0.4\linewidth]{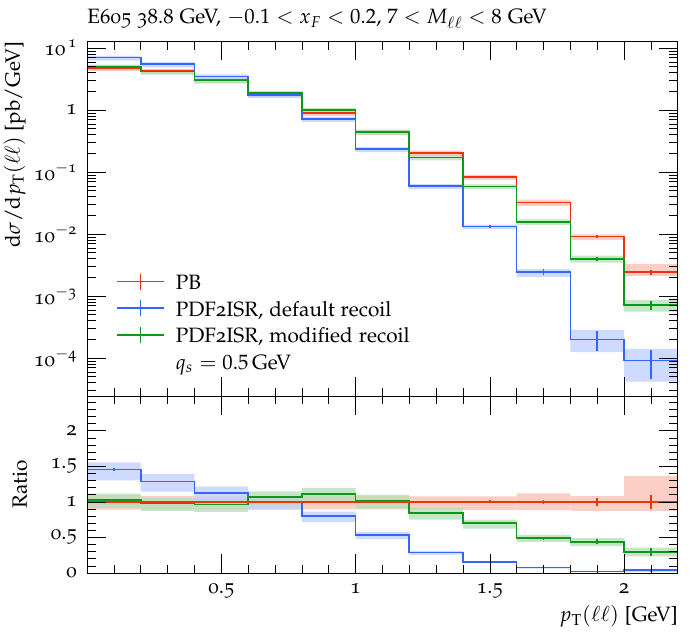}
\includegraphics[width=0.4\linewidth]{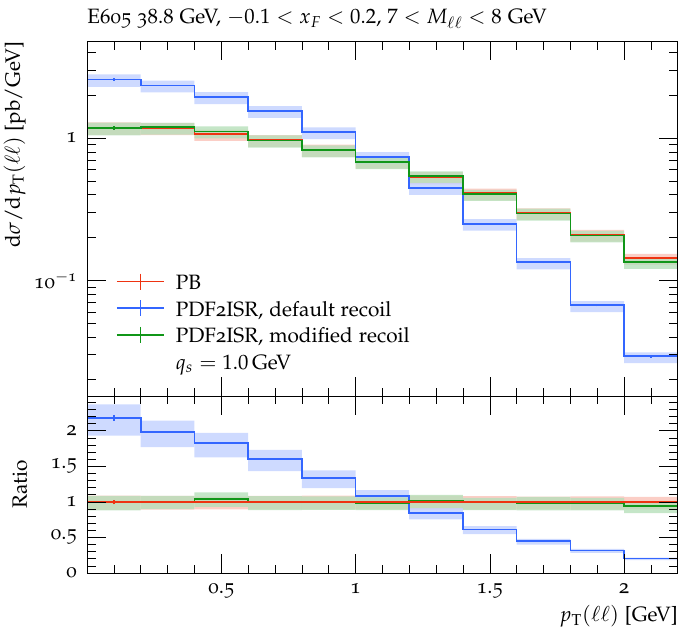}
\cprotect\caption {\small Comparison of \ptll\  for the DY pairs with invariant mass $7 \le m(\ell\ell) \le 8~\text{GeV}$ produced at $\sqrts=38.8$~\GeV\ for two different intrinsic-\kt\ widths in \pythiaPB\ and \PB\ (\cascade ): $\qs = 0.5$~GeV (left) and $\qs = 1.0$~GeV (right).
For comparison the prediction of  \pythiaPB\ with the default recoil scheme is shown.
The uncertainty bands show the 7-point theoretical uncertainties.}
\label{fig:pythia-PB-comp-ktonly}
\end{figure}

In order to perform a direct comparison with \PB, we have generated \PB-TMD distributions
with perturbative evolution, but including only intrinsic $\kt$ for the transverse momentum,
as well as distributions without any intrinsic $\kt$. We then compare predictions for DY
production obtained with \PB\ as implemented in \cascade\ with the corresponding ones
obtained with \pythiaPB, applying the same collinear PDF and the same NLO partonic
calculation.

Fig.~\ref{fig:pythia-PB-comp-ktonly} shows the DY transverse-momentum
distribution for the lowest mass bin, $7 \le m(\ell\ell) \le 8~\text{GeV}$, for the case
of intrinsic $\kt$ only, at a low collision energy of~$\sqrts~=~38.8~\text{GeV}$.
(Please note that the difference at $\ptll \geq 1$~GeV for small $\qs$ arises from
matching to the NLO contribution.) 
For comparison, the prediction of \pythiaPB\ with the default recoil scheme is shown, clearly illustrating the inconsistent treatment.
The discrepancy between the default scheme and \PB\
increases with the value of $\qs$, as expected from Fig.~\ref{fig:pythia-kt}.

\begin{figure}[htbp]
\centering
\includegraphics[width=0.4\linewidth]{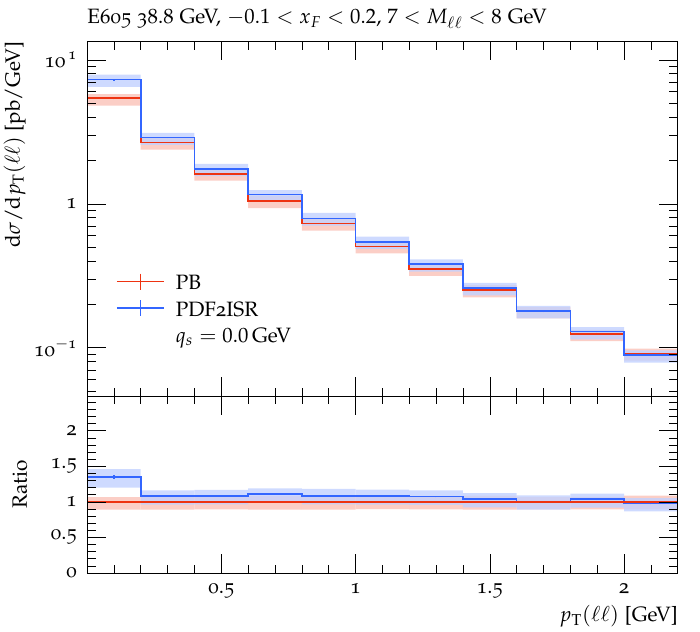}
\includegraphics[width=0.4\linewidth]{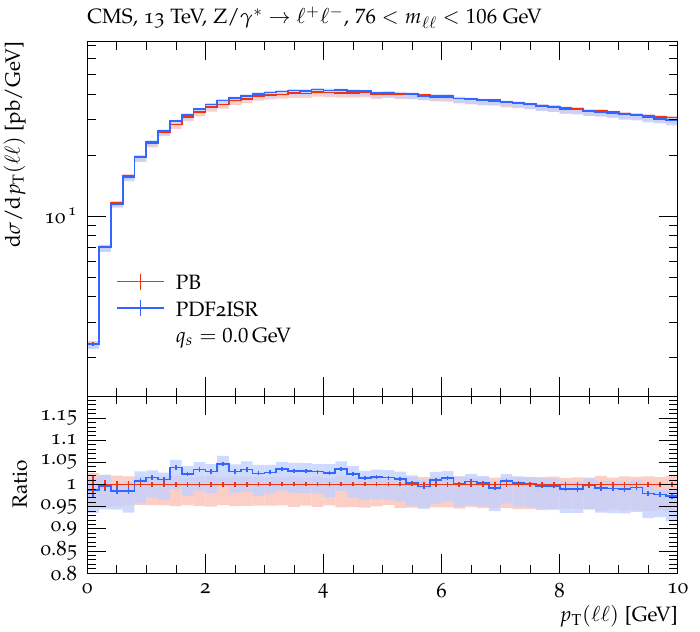}
\cprotect\caption {\small Comparison between \PB\ (\cascade ) and \pythiaPB\ predictions obtained with parton shower and TMD evolution (without intrinsic-\kt ) for \ptll\ of DY pairs with invariant mass $7 \le m(\ell\ell) \le 8~\text{GeV}$ produced at $\sqrts=38.8$~\GeV\ (left) and for DY pairs in the $Z$-peak region produced at $\sqrts=13$~\TeV\ (right).
The uncertainty bands show the 7-point theoretical uncertainties.}
\label{fig:pythia-PB-comp-nokt}
\end{figure}
These comparisons also demonstrate that intrinsic $\kt$ alone only provides a baseline
smearing of the transverse-momentum spectrum. The characteristic shape and normalisation
of the measured \ptll\ distribution, even at the lowest transverse momenta, require the
accumulation of transverse momentum through multiple soft-gluon emissions during the
initial-state evolution. 
Later in this section, we will explicitly verify that including ISR together with intrinsic-\kt\ describes the measured low-\pt\ spectra accurately, underscoring that intrinsic transverse momentum cannot substitute soft-gluon radiation.

As a further test, in Fig.~\ref{fig:pythia-PB-comp-nokt} we compare
predictions of \pythiaPB\ with \PB\ when only the parton shower and TMD evolution are
included, while intrinsic $\kt$ is switched off. The comparison is shown for two cases:
DY pairs with invariant mass $7 \le m(\ell\ell) \le 8~\text{GeV}$ produced at
$\sqrts = 38.8~\text{GeV}$, and DY pairs in the $Z$-peak region produced at
$\sqrts = 13~\text{TeV}$. The residual differences are attributed to boost effects,
since the calculation of $\kt$ is performed in different reference frames in \pythiaPB\
and \PB, as discussed in Ref.~\cite{Jung:2025mtd}.

In the forward \PB-TMD evolution, the intrinsic-transverse-momentum distribution enters
via the boundary condition in Eq.~(\ref{TMD_A0}),
\begin{equation}
x{\cal A}(x,{\bf k}_0,\mu_0^2)
= xf_0(x,\mu_0^2)\,G(\sigma,\kt)\,,
\end{equation}
with $G(\sigma,\kt) = \exp(-\kt^2 / 2\sigma^2)/(2\pi\sigma^2)$ a Gaussian distribution.

The evolution equation for the TMD ${\cal A}(x,{\bf k},\mu^2)$ at scale  $\mu$ is given by:
\begin{align}
\MoveEqLeft
x\mathcal{A}_a(x,\mathbf{k},\mu^2)
=
\Delta_a^S(\mu^2,\mu_0^2)\,
x\mathcal{A}_a(x,\mathbf{k},\mu_0^2)
\nonumber\\
&\hspace{-0.05cm}
+\sum_b
\int_{\mu_0^2}^{\mu^2}
\frac{d^2\mathbf{q}'}{\pi q'^2}\,
\Delta_a^S(\mu^2,q'^2)
\int_x^{z_M} dz\,
P_{ab}^{(R)}(\alpha_s,z)\,
\frac{x}{z}
\mathcal{A}_b
\left(
\frac{x}{z},
\mathbf{k}+(1-z)\mathbf{q}',
q'^2
\right) .
\label{eq:tmd_evolution}
\end{align}
where $\Delta^S_a$ denotes the same PB Sudakov form factor as in Eq.~(\ref{sud-def}).  Since, in \PB , the transverse momentum of the propagating parton \kt\ does not enter the evolution equation, but is reconstructed from the recoil momenta ${\bf q}_i$, the final transverse momentum is given by 
\begin{equation}
{\bf k} = \sum_i {\bf q}_i + {\bf k}_0 \, ,
\end{equation}
where ${\bf q}_i$ are the transverse momentum vectors of the emitted partons and ${\bf k}_0$ is the transverse momentum from the intrinsic-\kt\ distribution. 

In contrast, the current backward implementation in \pythiaPB\ constructs a parton ladder from the hard scale $\mu$ down to the starting scale $\mu_0$, boosting and rotating the whole system at each step such that the initiating partons are collinear. Each rotation and boost depends on the \kt\ of the corresponding emission, and therefore the final transverse momentum vector $\bf k$ depends on the transverse momenta of individual emissions (in contrast to \PB ). Consequently, the fact that the intrinsic-\kt\ distribution is included only after the full backward parton shower has been performed becomes relevant.

\begin{figure}[!b]
\centering
\includegraphics[width=0.4\linewidth]{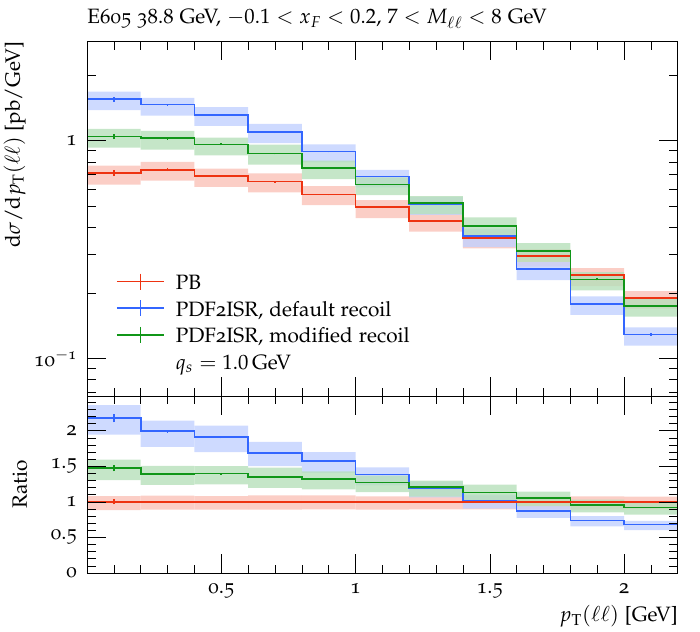}
\includegraphics[width=0.4\linewidth]{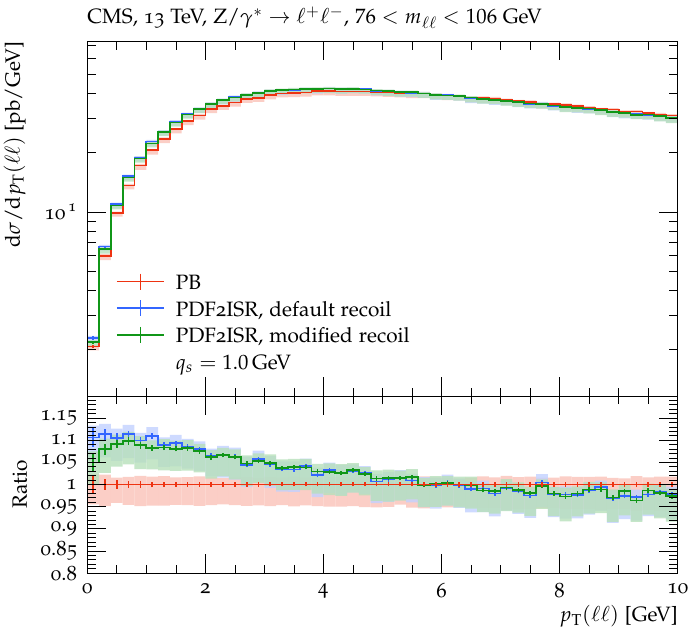}
\cprotect\caption {\small Comparison between \PB\ (\cascade ) and \pythiaPB\ predictions obtained with an intrinsic-\kt\ width of $\qs = 1~\text{GeV}$ and with parton shower and TMD evolution for \ptll\ of DY pairs with invariant mass $7 \le m(\ell\ell) \le 8~\text{GeV}$ produced at $\sqrts=38.8$~\GeV\ (left) and for DY pairs in the $Z$-peak region produced at $\sqrts=13$~\TeV\ (right).
For comparison the prediction of  \pythiaPB\ with the default recoil scheme is also shown.
The uncertainty bands show the 7-point theoretical uncertainties. Please note the different scales in the left and right ratio plots.}
\label{fig:pythia-PB-comp}
\end{figure}

Fig.~\ref{fig:pythia-PB-comp} shows a comparison between the \PB\ and \pythiaPB\ including an intrinsic-\kt\ distribution with $\qs = 1 $ \GeV\ and TMD evolution for \PB . In contrast to the case where only evolution was included, without an intrinsic-\kt\ distribution (see Fig.~\ref{fig:pythia-PB-comp-nokt}), significant differences are observed in the very small \ptll\ region. These differences are smaller for larger \sqrts\ and larger \mdy , where the role of intrinsic-\kt\ is smaller, and the evolution of soft gluons plays a larger role. For illustration, the predictions using the  default \pythia\ recoil scheme are also shown, demonstrating the significant effect at small  \sqrts\ and small  \mdy .

It is important to note that these differences are not a technical issue of implementation but are a consequence of the method used to reconstruct the kinematics in a backward-evolution approach like \pythia . The observed differences are tied to the kinematic reconstruction procedure of the \pythia\ backward-evolution shower.
In \herwig\ a different method of kinematic reconstruction is used, avoiding rotations and boosts for individual emissions~\cite{Herwig-shower-Plaetzer-2026}. It would therefore be interesting to study a \pdfisr implementation in \herwig~ as well.

\subsection{
 \pythiaPB\ predictions for Drell--Yan production 
  at different \texorpdfstring{$\bm{\sqrts}$}{sqrts}
}

In this section we describe the determination of \qs\ from measurements covering different \mdy\ and \sqrts\ ranges using the \pythiaPB\ approach described above. We use Drell--Yan production measurements (as in Ref.~\cite{Bubanja:2024puv}) at  $\sqrts = 38.8~\GeV$~\cite{Moreno:1990sf}, $\sqrts = 200~\GeV$~\cite{Aidala:2018ajl},  $\sqrts = 1.8~\TeV$~\cite{D0:1999jba}, $\sqrts = 1.96~\TeV$~\cite{CDF:2012brb}, $\sqrts = 8~\TeV$~\cite{Aad:2015auj}, and $\sqrts = 13~\TeV$~\cite{CMS:2022ubq}.
Uncorrelated and correlated systematic uncertainties of the measurements (where available)  are treated using the full covariance matrix provided in HEPData~\cite{hepdataplot}.
We calculate $\chi^2$ as a function of \qs\ for each experiment, and determine the best \qs\ value from the minimum of the $\chi^2$ distribution.\footnote{For the LHC and Tevatron energies, the \PZ -peak region is used, with the emphasis placed on the low transverse-momentum region (around or below the peak of the distribution). This region corresponds to the relevant non-perturbative domain in which the ISR simulated by \pythiaPB\  has the largest impact.}

\begin{table}[htp]
\caption{\small Goodness of the fits and extracted \qs\ ranges as a function of the centre-of-mass energy \sqrts , obtained by comparing \protect\pythiaPB\ predictions with data at different energies~\protect\cite{CMS:2022ubq,Aad:2015auj,CDF:2012brb,D0:1999jba,Aidala:2018ajl,Moreno:1990sf}.}

\begin{center}
\begin{tabular}{l|c|c|c}
Experiment                               & $\sqrt{s}$ &     \qs\ (GeV)      & $\chi^2/\text{n.d.f}$  \\ \hline
CMS~\cite{CMS:2022ubq}       & 13 \TeV    &       $1.7 - 1.9$         &       1.24         \\
CDF~\cite{CDF:2012brb}         & 1.96 \TeV &      $1.4 - 1.6$          &      0.40       \\
D0~\cite{D0:1999jba}               & 1.8 \TeV   &       $0.9 - 1.5$         &     0.50         \\
 PHENIX~\cite{Aidala:2018ajl} & 200 \GeV &       $0.9 - 1.2$         &      1.47        \\
E605~\cite{Moreno:1990sf}     & 38 \GeV    &       $0.9 - 1.0$         &      1.00       
\end{tabular}
\end{center}
\label{Tab:chi2}
\end{table}%

\begin{figure}[!t]
\centering
\includegraphics[width=0.8\linewidth]{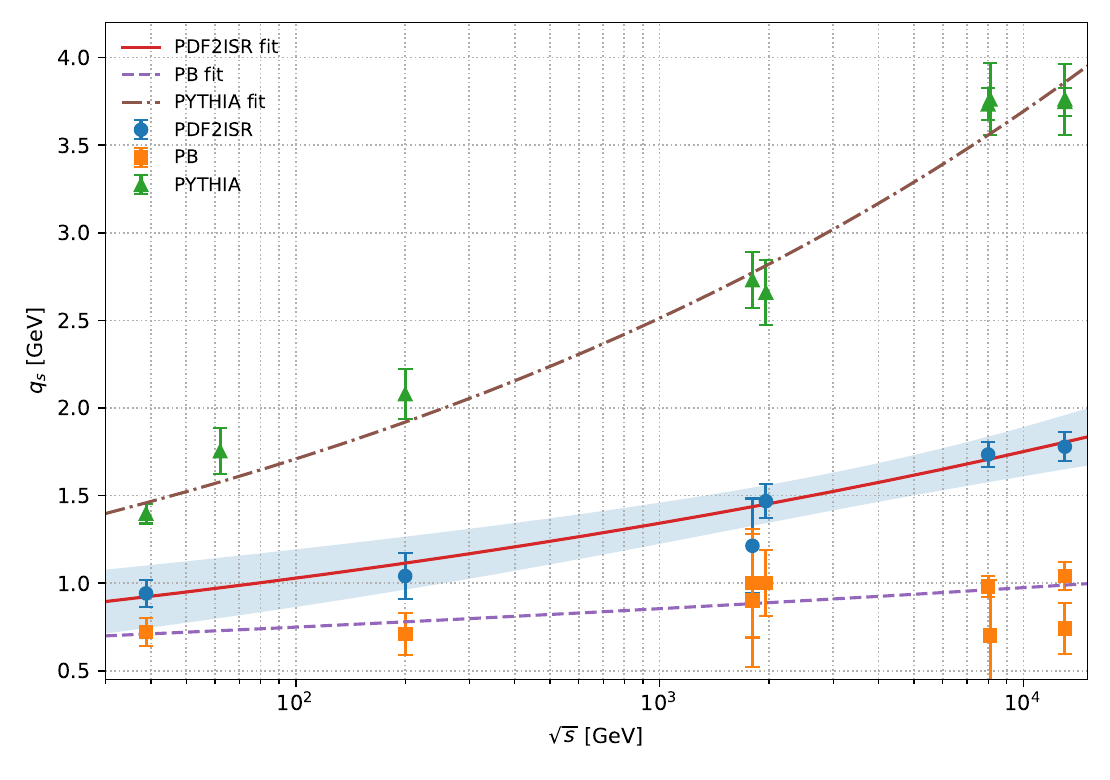}
\cprotect\caption {\small 
Values of \( \qs \) extracted from the comparison of \pythiaPB\ predictions with measurements at $\sqrts = 38.8~\GeV$~\cite{Moreno:1990sf}, $\sqrts = 200~\GeV$~\cite{Aidala:2018ajl}, $\sqrts = 1.8~\TeV$~\cite{D0:1999jba}, $\sqrts = 1.96~\TeV$~\cite{CDF:2012brb}, $\sqrts = 8~\TeV$~\cite{Aad:2015auj}  and $\sqrts = 13~\TeV$~\cite{CMS:2022ubq} as a function of \( \sqrts \), with uncertainties defined by \( \Delta \chi^2 = 1 \). The  shaded band corresponds to the 95\% confidence level interval of the fit. For comparison, the curves from standard \pythia~\cite{CMS:2024goo} and \PB~\cite{Bubanja:2024crf} are also shown. Please note that $\qs = \sqrt{2} \sigma$ is used for \pythia . }
\label{fig:DYchi2}
\end{figure}

The $\chi^2$ values obtained from the different measurements are given  in Tab.~\ref{Tab:chi2} together with the range obtained for  \qs\ defined by $\Delta \chi^2 = 1$. 
The  values of $\qs$ extracted from the individual fits, together with their uncertainty ranges,  are displayed in Fig.~\ref{fig:DYchi2} as a function of \sqrts . For comparison, the curve obtained from \PB\ in Ref.~\cite{Bubanja:2024crf} and the one obtained from \pythia\ in Ref.~\cite{CMS:2024goo} are also shown.  
A fit to the data points, together with the 95\% CL band, is shown to guide the eye. 
Compared to \PB , we observe a difference in  the width, \qs , of the Gaussian distribution for intrinsic-\kt ,  as well as a slightly steeper dependence on \sqrts . The  energy dependence of \qs\ obtained from \pythia\  shows a much steeper behaviour, which originates from the recoil treatment (as discussed above) and more importantly from the treatment of soft gluons as described in Ref.~\cite{Bubanja:2024puv}.

The discrepancy between the \qs\ values obtained from \pythiaPB\ and \PB\ reflects a deeper methodological distinction. Although the evolution parameters in the two approaches are the same, the remaining difference arises from the frame in which the transverse momentum of the initial-state partons is calculated, as noted in \cite{Jung:2025mtd}. As a consequence, the intrinsic-\kt\ distribution is affected when added at the end of the evolution. The observed differences are therefore not solely due to boosts, but are tied more generally to the event-level kinematic reconstruction procedure of the backward-evolution shower.

Although similar low-\pt\ spectra can be obtained after tuning, this does not imply that the underlying microscopic dynamics are identical. In the \PB -TMD approach, transverse momentum is generated dynamically during forward evolution and enters consistently into the Sudakov evolution. In contrast, within the backward-evolution framework of \pythiaPB , part of the observed transverse momentum is effectively absorbed into the fitted intrinsic-\kt\ distribution together with kinematic recoil effects.

In the following, we compare the data with \pythiaPB\ predictions obtained using the corresponding optimal value of $\qs$ for each data set (for comparison we also show the prediction using the default recoil treatment).
\begin{figure}[!t]
\centering
\includegraphics[width=0.32\linewidth]{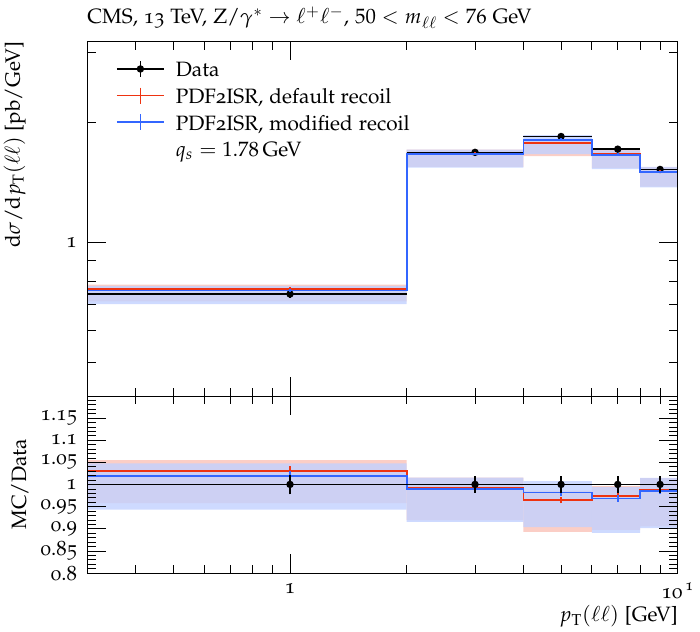}
\includegraphics[width=0.32\linewidth]{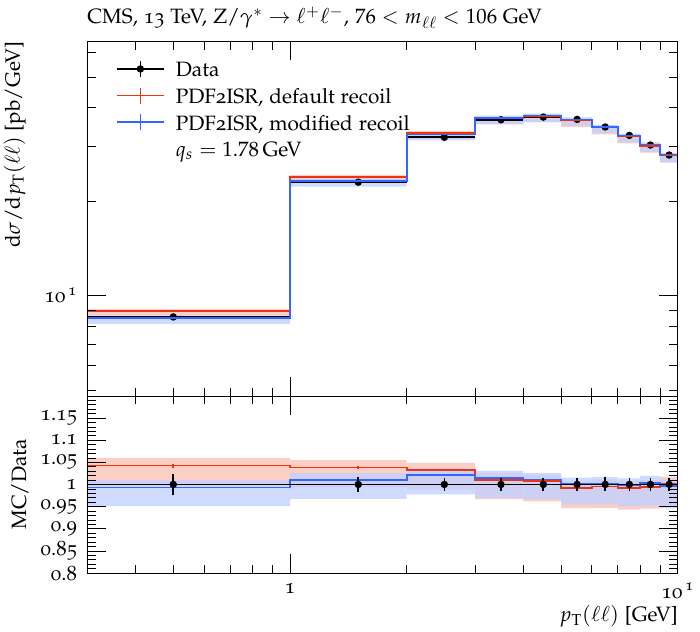}
\includegraphics[width=0.32\linewidth]{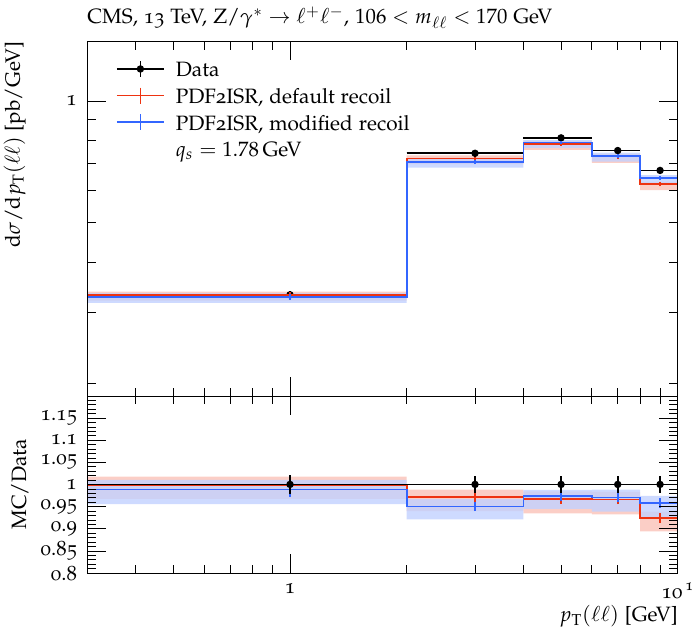}
\cprotect\caption {\small Transverse-momentum distributions of DY lepton pairs at $\sqrts = 13$~\TeV obtained with the \pythiaPB approach using PB-NLO-2018-Set2 compared to measurements~\cite{CMS:2022ubq} in three invariant-mass bins, as indicated above the histograms. 
The uncertainty bands show the 7-point theoretical uncertainties.
The predictions using the default recoil treatment are also shown for comparison.}
\label{fig:DY13TeV}
\end{figure}
Fig.~\ref{fig:DY13TeV} shows the predictions of the \pythiaPB\ approach compared with CMS measurements~\cite{CMS:2022ubq} at the highest centre-of-mass energy, $\sqrts = 13$~TeV, for the three invariant-mass bins indicated in the figure, focusing on the low-\ptll\ region. 

\begin{figure}[!b]
\centering
\includegraphics[width=0.32\linewidth]{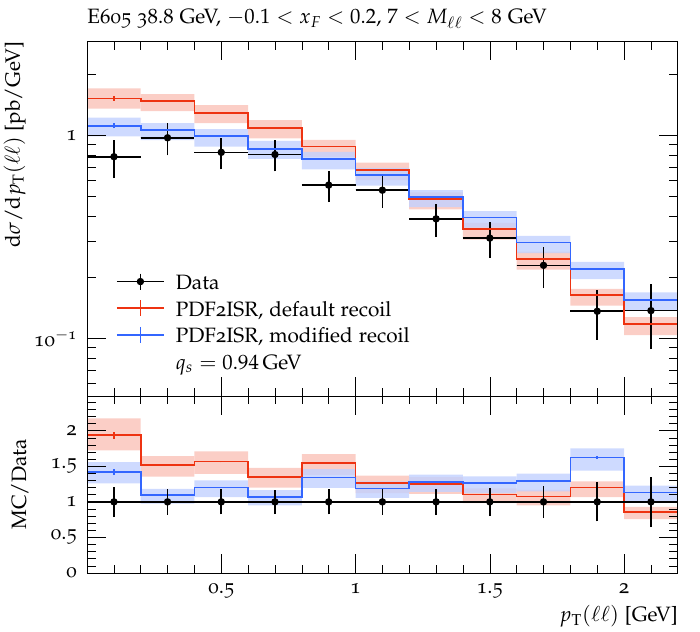}
\includegraphics[width=0.32\linewidth]{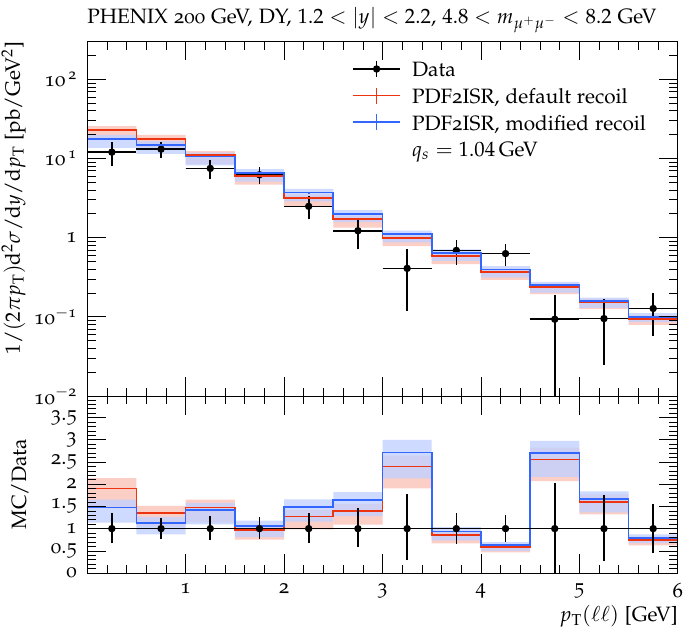}
\includegraphics[width=0.32\linewidth]{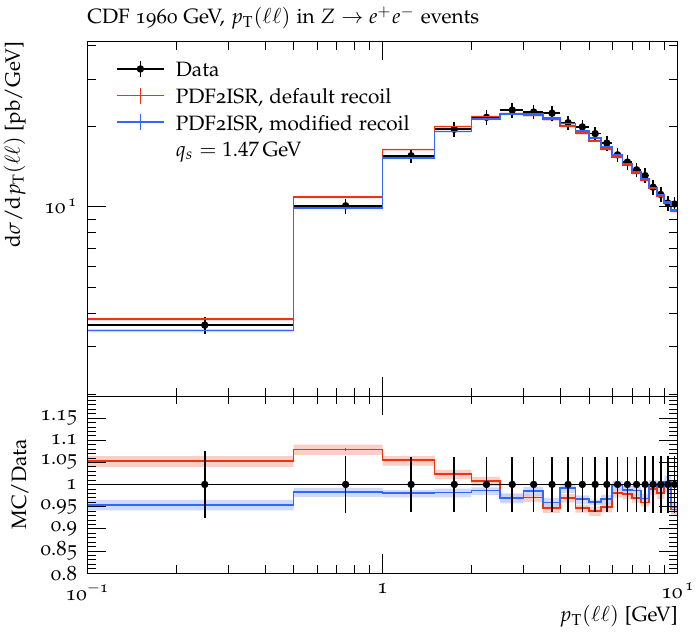}
\cprotect\caption {\small Transverse-momentum distributions of DY lepton pairs at $\sqrts = 38.8$~\GeV\ (left), $\sqrts = 200$~\GeV\ (middle), and $\sqrts = 1.96$~\TeV\ (right), obtained with the \pythiaPB approach using PB-NLO-2018-Set2 
compared to measurements~\cite{Moreno:1990sf,Aidala:2018ajl,CDF:2012brb}. The invariant-mass ranges of the DY pairs are indicated above the histograms, and the uncertainty bands show the 7-point theoretical uncertainties. The predictions using the default recoil treatment are also shown for comparison.
}
\label{fig:DYlowE}
\end{figure}
In Fig.~\ref{fig:DYlowE}, the \pythiaPB\  predictions are compared with measurements performed at much lower energies of  $\sqrts = 38.8$~\GeV~\cite{Moreno:1990sf} and $\sqrts = 200$~\GeV ~\cite{Aidala:2018ajl}. In addition, a comparison with Tevatron data in the $Z$-peak region at $\sqrts = 1.96~\TeV$ is shown~\cite{CDF:2012brb}.

From these comparisons we conclude that, with the improved recoil treatment and the fitted values of \qs, \pythiaPB\ provides a good description of the measured low-\ptll\ spectra over a wide range of centre-of-mass energies. This agreement indicates that the interplay between intrinsic \kt\ and soft-gluon emissions is consistently captured within the \pythiaPB\ initial-state shower.

The comparisons in this section have been performed at NLO, applying NLO splitting functions in the initial-state parton shower, as described in Ref.~\cite{Jung:2025mtd}. Since the small-\ptll\ region receives contributions from both intrinsic \kt\ and soft-gluon emissions in the parton shower, higher-order splitting functions are expected to modify the relative importance of the intrinsic-\kt\ contribution.

\section{%
  The non-perturbative region in
  \texorpdfstring{$\bm{\alpha_{\rm s}}$}{alpha\_s}%
}

The infrared behaviour of the strong coupling \as\ has been addressed in early works~\cite{Cornwall:1981zr} on a dynamical, running gluon mass~\cite{Oliveira:2010xc,Bicudo:2014cqa,Aguilar:2014tka,Deur:2023dzc}, as well as in the pioneering work~\cite{Shirkov:1997wi} on analytic continuation of the strong coupling~\cite{Solovtsov:1999in,Bakulev:2005gw,Kotikov:2022vnx,Kotikov:2022sos,Kotikov:2023meh}.  A recent review on the strong coupling is given in Ref.~\cite{Deur:2025rjo}. Latest developments have been presented in Ref.~\cite{PoS-QCDExtremes2025}. A discussion of \as\ in the infrared region in the context of TMDs is given in Ref.~\cite{Simonelli:2025kga}.

The low-\ptll\ distributions in DY production are also sensitive to the non-perturbative region of \as .
In \PBset~Set2, the transverse momentum of the emitted parton, \qt , is used as the scale in \as .\ 
Since soft gluons are explicitly taken into account down to $q_0 = 0.01~\GeV$, the emitted-gluon transverse momentum, \qt , is not restricted. The coupling \as\ must therefore be modified to avoid the region below the Landau pole, $\qt < \Lambda_{\rm QCD}$.

The purpose of this study is not to determine a preferred infrared form of \as , but to quantify how sensitive low-\pt\  DY observables are to the transition region between perturbative and non-perturbative scales when soft ISR is treated consistently. The sensitivity is most pronounced in the first few \GeV\ of the dilepton transverse-momentum spectrum at the highest collider energies considered in this study.

\begin{figure}[!b]
\centering
\includegraphics[width=0.5\linewidth]{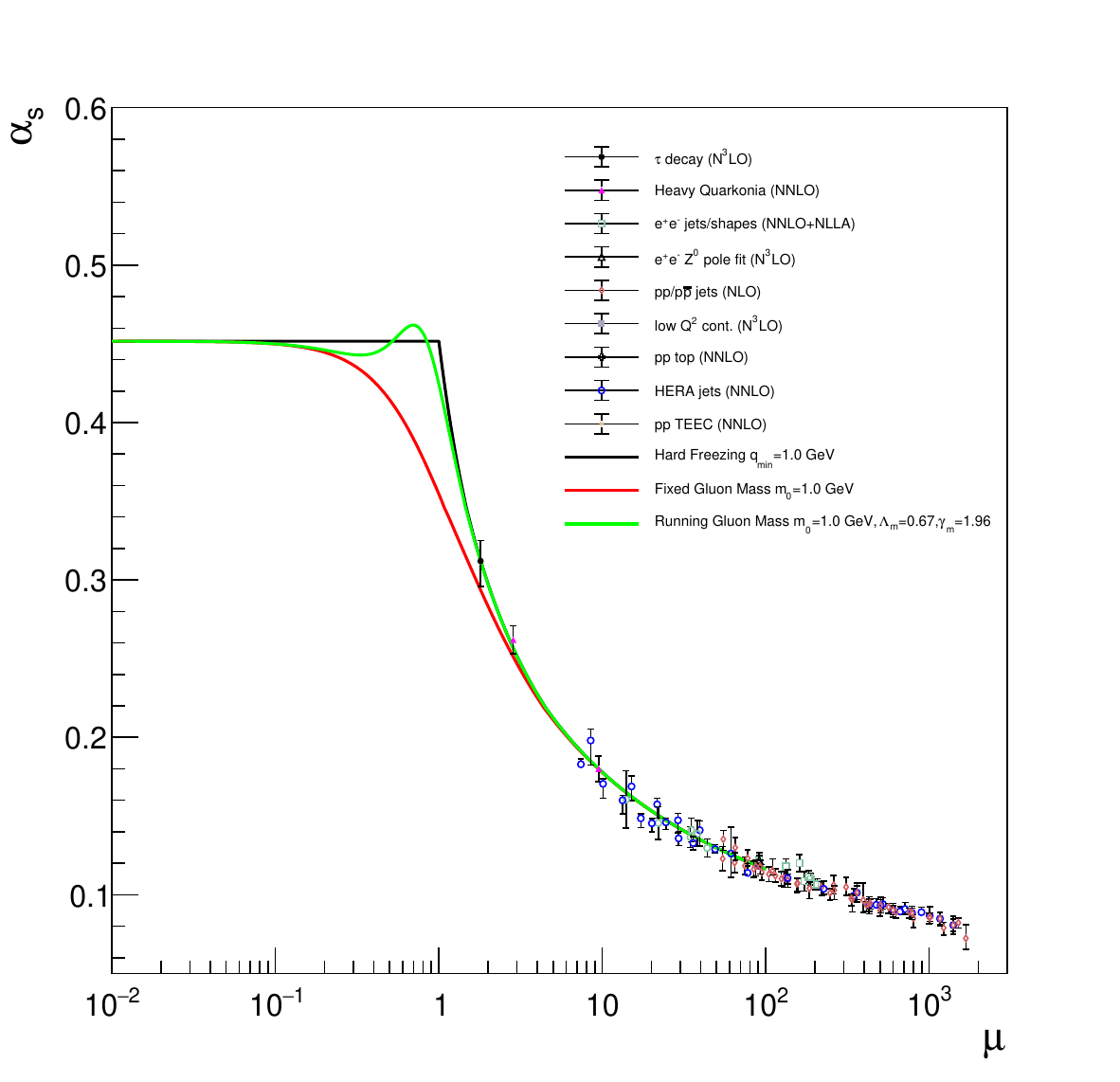}
\cprotect\caption {\small \as\ as a function of the scale $\mu$, with experimental measurements~\cite{ATLAS:2023tgo,FlavourLatticeAveragingGroupFLAG:2021npn,ParticleDataGroup:2022pth,Britzger:2019kkb,Narison:2018xbj,Aaboud:2018vdd,Boito:2018yvl,ATLAS:2017qir,H1:2017bml,Khachatryan:2016mlc,Klijnsma:2017eqp,CMS:2014qtp,CMS:2014mna,CMS:2013vbb,Schieck:2012mp,D0:2012xif,OPAL:2011aa,D0:2009wsr,Bethke:2009ehn,Dissertori:2009ik,Glasman:2005ik} and predictions at NLO, for Hard Freezing (Option A), Fixed Gluon Mass  (Option B), and Running Gluon Mass  (Option C), with the parameters given in the text.}
\label{fig:alphas}
\end{figure}

In the following, we consider three different prescriptions:
\begin{itemize}
\item {\bf A: Hard Freezing}:  As implemented in the default \PBset~Set2~\cite{Martinez:2018jxt}, a freezing cutoff $ q_{\rm min}$ is introduced, such that:
\begin{equation}
\as = \as(\max(q_{\rm min} ^2 ,\mu^2)) .
\end{equation}
The default value of the scale at which the transition from the perturbative to the non-perturbative treatment of \as\ is realised is taken to be $q_{\rm min} =1~\GeV$.
\item {\bf B: Fixed Gluon Mass:} A fixed ``gluon mass''~\cite{Braun:1994mw} $ m_0  = 1 ~\GeV$  can be introduced to obtain a smooth transition: 
\begin{equation}
\as = \as(m_0^2 + \mu^2) .
\end{equation}
\item {\bf C: Running Gluon Mass:} A ``running gluon mass''~\cite{Aguilar:2014tka}, which itself depends on the scale $\mu$, can be introduced:
\begin{equation}
\as = \as\left(m(\mu^2)^2 + \mu^2\right) ,
\end{equation}
where
\begin{equation}
m^2(\mu^2) = \frac{m_0^2}{1 + (\mu^2/\Lambda_m^2)^\gamma},
\end{equation}
with $m_0 = 1 ~\GeV$, $\Lambda_m = 0.67~\GeV$, and $\gamma = 1.96$.
To facilitate the comparison, the parameters $m_0$, $\Lambda_m$, and $\gamma$ are chosen to best reproduce option~A.
\end{itemize}

\begin{figure}[!b]
\centering
        \includegraphics[width=0.32\linewidth]{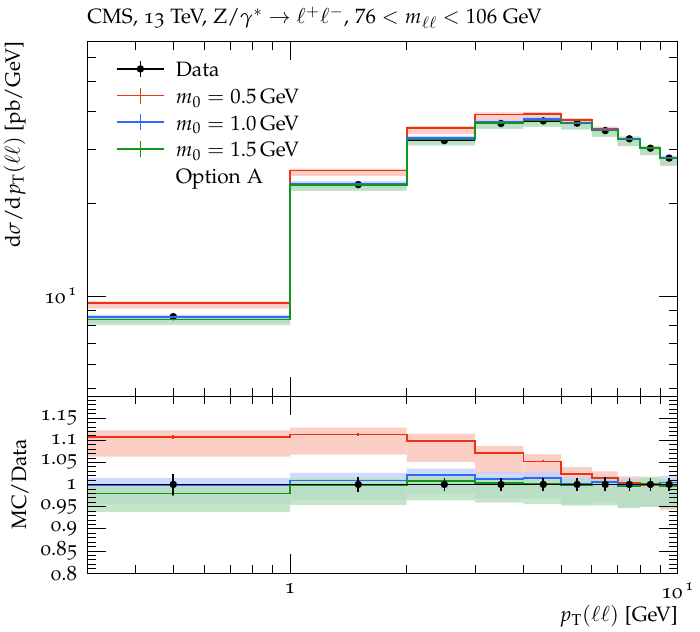}
        \includegraphics[width=0.32\linewidth]{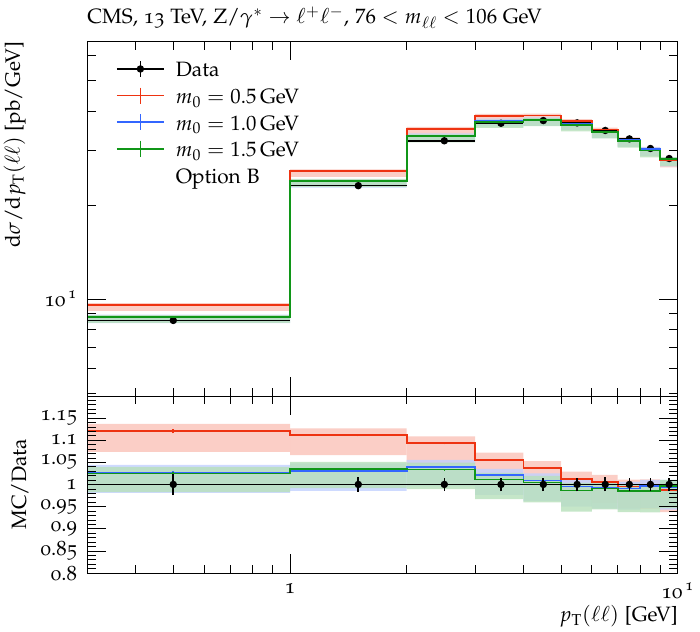}
        \includegraphics[width=0.32\linewidth]{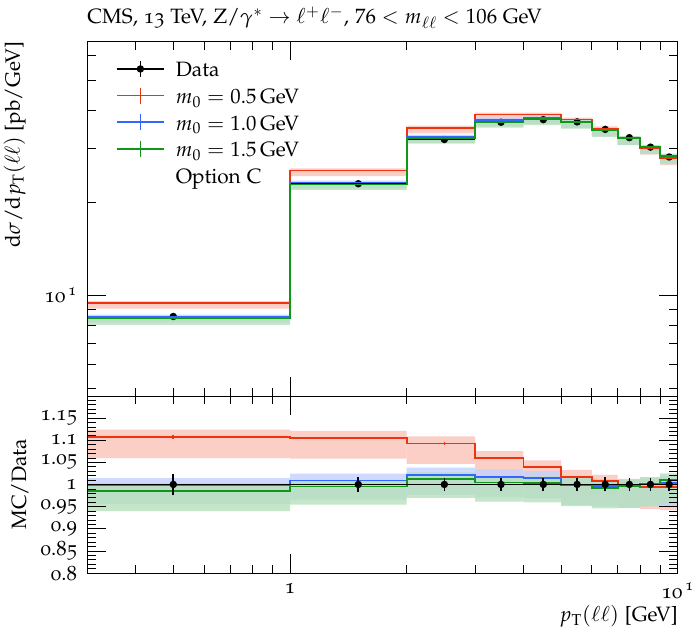}
\cprotect\caption {\small Transverse-momentum distributions of DY lepton pairs at $\sqrts = 13~\TeV$, obtained with the \pythiaPB approach, using three different ways of taming \as\ and three values of the $m_0$ parameter, compared to measurements~\cite{CMS:2022ubq}. The uncertainty bands show the 7-point theoretical uncertainties. }
\label{fig:HE_as}
\end{figure}

In Fig.~\ref{fig:alphas}, we show a summary of the measurements of \as~\cite{ATLAS:2023tgo,FlavourLatticeAveragingGroupFLAG:2021npn,ParticleDataGroup:2022pth,Britzger:2019kkb,Narison:2018xbj,Aaboud:2018vdd,Boito:2018yvl,ATLAS:2017qir,H1:2017bml,Khachatryan:2016mlc,Klijnsma:2017eqp,CMS:2014qtp,CMS:2014mna,CMS:2013vbb,Schieck:2012mp,D0:2012xif,OPAL:2011aa,D0:2009wsr,Bethke:2009ehn,Dissertori:2009ik,Glasman:2005ik} and the different options for calculating  \as\  at NLO in the small-\qt\ region.

Fig.~\ref{fig:HE_as} shows predictions obtained with \pythiaPB using three different ways of taming \as\ in the small-\ptll\ region, compared to CMS measurements in the \PZ -peak region at $\sqrts=13\,\mathrm{TeV}$~\cite{CMS:2022ubq}. 
 The values $q_{\rm min}=m_0 = 0.5, 1.0, 1.5$~\GeV\ are shown as illustrative variations of the effective low-scale coupling in the shower.
As expected, smaller values of the infrared scale (smaller $m_0$) lead to a larger coupling in the low-scale region. This enhances soft initial-state radiation and produces a visible excess in the lowest-\ptll\ bins, which is observed consistently for the  $m_0=0.5\,\GeV$ choice in all three prescriptions. In contrast to that, the $1.0$ and $1.5\,\GeV$ choices give similar spectra, indicating that once the infrared coupling is sufficiently moderated, the final \ptll\ distribution is only weakly affected by further changes of the infrared scale. 
\begin{figure}[!t]
\centering
         \includegraphics[width=0.32\linewidth]{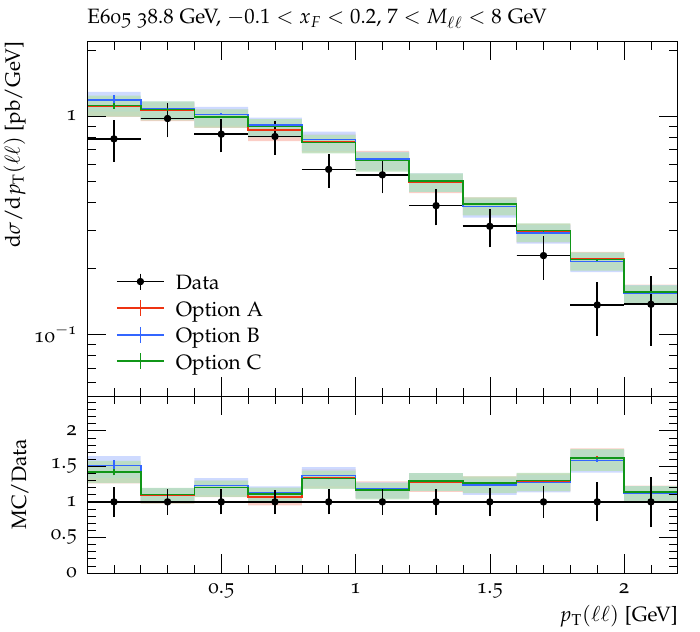}
         \includegraphics[width=0.32\linewidth]{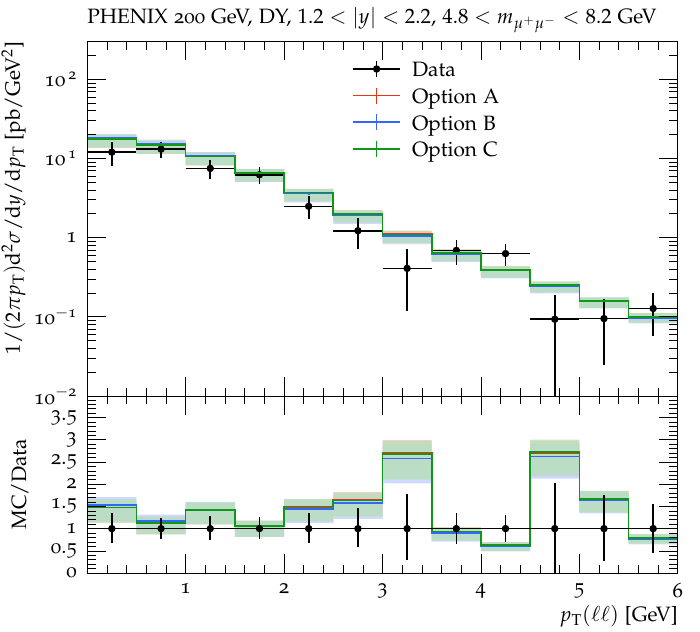}
         \includegraphics[width=0.32\linewidth]{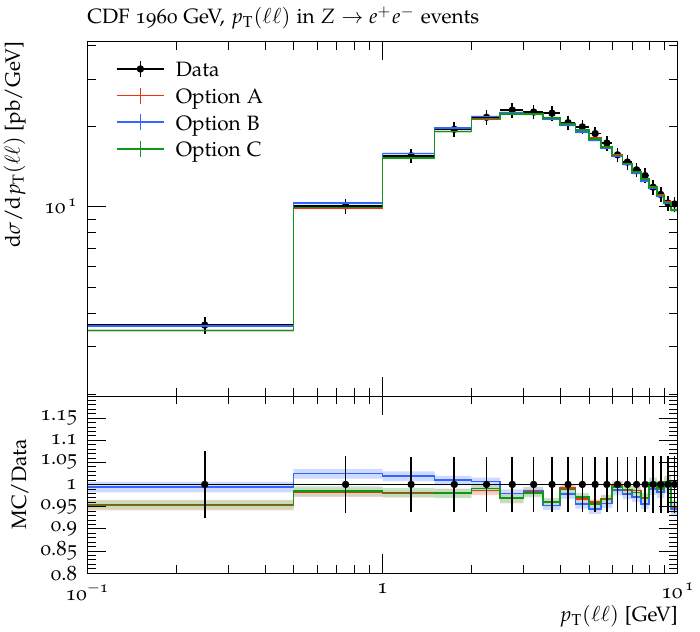}
\cprotect\caption {\small Transverse-momentum distributions of DY lepton pairs at $\sqrts=38.8$~\GeV\ (left), $\sqrts~=~200$~\GeV\ (middle), and 
$\sqrts~=~1.96$~\TeV\ (right), obtained with the \pythiaPB approach, using three different ways of taming \as\ with $m_0 = 1$~\GeV, compared to measurements~\cite{Moreno:1990sf,Aidala:2018ajl,CDF:2012brb}. The invariant-mass ranges of the DY pairs are indicated above the histograms. The uncertainty bands show the 7-point theoretical uncertainties.}
\label{fig:LE_as}
\end{figure}

The detailed functional form for the infrared continuation of \as\ leads to visible differences in the predictions, with the fixed-gluon-mass prescription exhibiting a larger deviation from the measurements relative to the hard-freezing and running-gluon-mass prescriptions. Although these deviations remain covered by the scale-uncertainty band, they are larger than the experimental uncertainties in the low-\ptll\ region.

\begin{tolerant}{8000}
We focus the comparison of the infrared functional forms on the nominal value $q_{\rm min}=m_0 =1.0 $~\GeV , which is consistent with the setup used in the construction of the underlying PB parton densities. In Fig.~\ref{fig:LE_as}, we compare \pythiaPB predictions using different treatments of \as\ to low-energy measurements of DY pairs with masses of a few \GeV , produced at $\sqrts~=~38.8~\GeV$~\cite{Moreno:1990sf} and 
$\sqrts~=~200~\GeV$~\cite{Aidala:2018ajl}, as well as to data in the $Z$-peak region obtained at $\sqrts=1.96~\TeV$~\cite{CDF:2012brb}. For the fixed-target and RHIC data, the differences between the three infrared prescriptions are small compared with the experimental uncertainties, and the data do not provide meaningful discrimination between the models. At Tevatron energies, although the differences remain covered by the uncertainty band, a somewhat larger prescription-dependent shape variation becomes visible in the low-\ptll\ region. 
A similar effect is seen at $\sqrts=13~\TeV$, indicating that soft-gluon emissions play a more important role at higher \sqrts\ than at lower energies.
\end{tolerant}

At this stage, we do not attempt to perform a fit of $m_0$ for the different options. Instead, we emphasize the main observation that the details of the transition from the perturbative to the non-perturbative regime of \as\ become visible in the description of low-\ptll\ DY spectra at Tevatron and LHC energies. This is of particular interest in the high-statistics $Z$-peak region, where the small-\ptll\ cross section becomes sensitive to the low-scale behaviour of the strong coupling.

\section{Conclusion}
We have applied the \pythiaPB approach to describe the non-perturbative contributions to the low-\pt\ region of the Drell--Yan lepton-pair spectrum. In particular, we considered the intrinsic transverse motion of partons and multiple soft-gluon emissions over a wide range of centre-of-mass energies, spanning from those covered by fixed-target experiments to those at the LHC.

The results show that \pythiaPB provides a description of the initial-state radiation contribution in the low-\pt\ DY region that is  comparable to that obtained with PB-TMD calculations across a broad range of DY invariant masses and collision energies.
To ensure a consistent comparison of the intrinsic partonic transverse momentum contribution in the two approaches, we performed a detailed investigation of how the treatment of parton momenta in the recoil-sharing step and the reference-frame boost affect the Drell--Yan \pt\ spectra. This study allowed us to disentangle technical effects from genuine physical effects that influence the non-perturbative contributions in the low-\pt\ DY distribution.
The remaining differences between the \pythia -based \pythiaPB and \PB -TMD approaches are primarily related to their different treatments of intrinsic transverse momentum and recoil kinematics, including effects induced by reference-frame rotations and boosts.
Despite these conceptual differences, the \pythiaPB approach provides a description of the experimental data that is comparable in quality to that of the PB-TMD approach.

A central outcome of this study is the clear identification of the complementary roles played by intrinsic transverse momentum and soft initial-state radiation. 
The fitted intrinsic-\kt\ distribution in \pythiaPB effectively absorbs not only genuinely non-perturbative transverse motion, but also part of the kinematic effects associated with recoil sharing and frame transformations in the backward-evolution algorithm.

We further find that, within the \pythiaPB approach based on a parton shower consistent with the underlying parton densities, low-\pt\
Drell--Yan predictions become sensitive to the treatment of \as\ in the transition region between perturbative and non-perturbative scales at the highest \sqrts , particularly in the first few GeV bins of the dilepton transverse-momentum spectrum.  

In summary,  the low-\pt\ Drell--Yan spectrum emerges from the interplay of intrinsic parton motion and soft-gluon radiation in the initial state. 
Both ingredients are essential for a quantitatively reliable description of the low-\ptll\ spectrum and must be treated consistently.
The comparison of the \pythiaPB\ predictions with the conventional \PB\ approach, as well as with experimental data, 
supports the phenomenological consistency of this physical picture  and underlines the potential of Drell--Yan \pt\  measurements to constrain non-perturbative QCD dynamics and the infrared behaviour of the strong coupling.
In this sense, low-\pt\ Drell--Yan production serves not only as a benchmark for TMD and parton-shower formalisms, but also as a potential probe of the transition region of the strong coupling, complementary to traditional extractions of \as\ based on event shapes and jet observables.

\vskip 0.5 cm
\begin{tolerant}{8000}
\noindent
{\bf Acknowledgments}
We are grateful to D. Britzger and  D. d'Enterria for providing the data points for the \as\ measurements.
We are grateful to A. Bagdatova, S. Baranov,  A. Lipatov, M. Malyshev, G. Lykasov and the other participants  of the WeeklyOfflineMeeting for many useful discussions during the past years. 
This article is part of a national scientific project that has received funding from the Montenegrin Ministry of Education, Science and Innovation. The authors also acknowledge the networking support from COST Action CA24159 (SHARP).
\end{tolerant}

\bibliographystyle{mybibstyle-new.bst}
\raggedright  
\input{DY-PDF2ISR-2026-06-09.bbl}

\end{document}

%% file: DY-PDF2ISR-2026-06-09.bbl
\providecommand{\href}[2]{#2}\begingroup\raggedright\endgroup